\documentclass[preprint]{imsart}

\usepackage{color}

\usepackage{geometry}                
\usepackage{graphicx}
\usepackage{epstopdf}
\usepackage{epsfig}
\usepackage{url}

\RequirePackage[OT1]{fontenc}
\RequirePackage{amsthm,amsmath,natbib}
\RequirePackage[colorlinks,citecolor=blue,urlcolor=blue]{hyperref}
\RequirePackage{hypernat}

\startlocaldefs
\numberwithin{equation}{section}
\theoremstyle{plain}

\endlocaldefs

\begin{document}

\begin{frontmatter}

\title{ Bayesian Nonparametric Variable Selection as an Exploratory Tool for Finding Genes that Matter}
\runtitle{}

\begin{aug}
\author{\fnms{Babak} \snm{Shahbaba}\ead[label=e1]{babaks@uci.edu} and \fnms{Wesley O.} \snm{Johnson}\ead[label=e2]{wjohnson@uci.edu}}

\runauthor{B. Shahbaba and W. O. Johnson}
\affiliation{University of California at Irvine}
\address{Department of Statistics, University of California at Irvine, CA, USA}

\end{aug}

\begin{abstract}
High-throughput scientific studies involving no clear a'priori hypothesis are common. For example, a large-scale genomic study of a disease may examine thousands of genes without hypothesizing that any specific gene is responsible for the disease. In these studies, the objective is to explore a large number of possible factors (e.g. genes) in order to identify a small number that will be considered in follow-up studies that tend to be more thorough and on smaller scales. For large-scale studies, we propose a nonparametric Bayesian approach based on random partition models. Our model thus divides the set of candidate factors into several subgroups according to their degrees of \emph{relevance}, or potential effect, in relation to the outcome of interest. The model allows for a latent rank to be assigned to each factor according to the overall potential importance of its corresponding group. The posterior expectation or mode of these ranks is used to set up a threshold for selecting potentially relevant factors. Using simulated data, we demonstrate that our approach could be quite effective in finding relevant genes compared to several alternative methods. We apply our model to two large-scale studies. The first study involves transcriptome analysis of infection by human cytomegalovirus (HCMV). The objective of the second study is to identify differentially expressed genes between two types of leukemia.

\end{abstract}
\begin{keyword}
\kwd{Random partition models}
\kwd{High throughput studies}
\kwd{Dirichlet process mixtures}
\kwd{Gene expression microarrays}
\end{keyword}

\end{frontmatter}
\setcounter{page}{0}

\baselineskip = 20pt

\eject

\section{Introduction}

High throughput studies are typically aimed at assessment of a large number of factors with respect to their relationship to an outcome of interest (e.g. disease status). The overall objective is to select a subset of factors that might be \emph{relevant} to the outcome. In this paper, we discuss two such studies as motivating examples. The objective of the first study, conducted by \cite{chan08}, is to identify genes that are differentially expressed after human cytomegalovirus (HCMV) infection. Infection by HCMV leads to morbidity and mortality in immunocompromised individuals, including AIDS and organ transplant patients. In this study, the expression levels of 12,626 genes were compared between six HCMV-infected (case group) and six mock-infected (control group) samples. The second study, conducted by \cite{armstrong02}, aims at identifying differentially expressed genes in two types of leukemia: acute myeloid leukemia (AML) and acute lymphoid leukemia (ALL). The data include the expression levels of 10,056 genes for 48 subjects (24 subjects in each group).

There are many papers that have approached this problem from the point of view of assessing ``statistical significance'' while controlling the family-wise error rate.  Here, we do not attempt to assess statistical significance, but we maintain the same goal of identifying factors that warrant further study.  Indeed, the method we propose, though model based, is regarded as exploratory data analysis where we search through a rather large number of genes in search of a subset of them that might be important or relevant.  High throughput studies can be regarded as a starting point to generate a small number of hypotheses that are worthy of further investigation with follow-up studies. Our approach is nonparametric and employs a latent random partition model based on Dirichlet Process mixtures (DPM) to divide genes into subgroups. Genes are ranked according to the overall potential importance of their corresponding groups. The posterior expectation of these ranks could be used to set up a threshold for selecting a relevant set.  Our method can be viewed in the context of variable selection where there could potentially be a very large number of terms in the model but where there is a belief in sparsity, which translates to parsimony.

Without loss of generality, we focus on the analysis of gene expression microarray data as a typical high throughput study. Gene expression studies deal with identifying genetic factors that are potentially important to an outcome of interest. Microarrays measure the expression levels for thousands of genes simultaneously. We denote the genes as $\mathcal{G}_{1}, \ldots, \mathcal{G}_{N}$. The outcome of interest is disease status (i.e., diseased vs. healthy), which is fixed in a case-control design.  The gene expression data consist responses on multiple diseased and healthy individuals for each gene, namely, for gene $i$ we observe $y_i = (y_{ijk}: j = 1,...,n_{ik}, \, k=0,1)$ where $k=0 $ corresponds to healthy and $k=1$ to diseased. Disease status is represented by a binary variable, $x_{ijk}$, which takes the value one if $k=1$ and zero otherwise.

Within the hypothesis testing framework, the usual statistical methods assume that for each gene, $\mathcal{G}_{i}$, there is a corresponding [null] hypothesis, $H_{i}$, stating that there is no change in gene expression between the two groups (diseased vs. healthy). Based on this assumption and the observed expression data, $y_{i}$, a simple test statistic $z_{i}$ is often computed for each gene such that the distribution of $Z_{i}$ is known under the null hypothesis. In general, larger values of $z_{i}$  provide stronger evidence of departure from $H_{i}$ and statistics above a certain cutoff are deemed significant, after adjustment to control the family-wise error rate or false discovery rate (FDR). \citep[See for example, ][]{hochberg88, hommel88, westfall93, benjamini95, storey04}.  As argued by several authors \citep[e.g., ][]{storey07, storey07b, sun09, guindani09}, methods that are based on $Z$-scores calculated for each test individually ignore information from other tests.

\cite{efron01} introduce the local false discovery rate (locFDR), which is the empirical Bayes version of the method proposed by \cite{benjamini95} for estimating the \emph{false discovery rate} (FDR).  The fully Bayesian version improves the performance of multiple significance testing by borrowing information across all tests when assessing the relative significance of each one of them.  See, for example, \cite{newton01,scott06,do05,muller06}. See \cite{scott10} for discussion regarding the comparison of fully Bayesian and empirical Bayesian adjustment for multiplicity.
In what follows, we present approaches based on the actual data and also based on summary $Z$ statistics for completeness.

\cite{storey07b} and \cite{storey07} proposed a related method called the \emph{optimal discovery procedure} (ODP), which is approximately equivalent to minimizing the \emph{missed discovery rate} for each fixed FDR. More recently, \cite{cao09} proposed a hierarchical Bayesian model, whose estimates are utilized in the ODP. We refer to this method as the Bayesian optimal discovery procedure (BODP). \cite{guindani09} (GMZ) showed that the ODP could be interpreted as an approximate Bayes rule under a semiparametric model. They proposed a Bayesian discovery procedure (BDP) that improves the approximation and allows for multiple shrinkage in clusters implied by a Dirichlet process mixture (DPM) model.  Using DP priors in the context of multiple hypothesis testing has been discussed by several other authors. See for example, \cite{gopalan93, dahl07, bogdan08}. Both ODP and BDP show improvement over some commonly used procedures such as SAM \citep[Significance Analysis of Microarrays, ][]{tusher01}, empirical Bayes, and local FDR.

In the remainder of the paper, we discuss our approach in details and compare its performance to several alternative methods. In Section 2, we present the basic models for the full data and reduced data. We also present the GMZ method and re-casts it for our purposes since it is related to ours and because of its success in comparison with other methods. Section 3 presents our data analyses followed by comparisons based on simulated data in section 4, sensitivity analysis in Section 5, and final conclusions in section 6.

\section{The Basic Model and Method}\label{sec:method}

\subsection{Model for the Full Data}\label{fullData}

Let $y_{ijk} $ be expression values and let $x_{ijk}$ denote the fixed values for group membership (healthy/diseased) as discussed above.  Denote the full data set as $y=\{y_{ijk}\}$. Then the basic model for these data is:
\begin{eqnarray}
y_{ijk} \mid  \alpha_{i}, \beta_{i} & \buildrel {ind} \over { \sim }& N(\alpha_{i}+\beta_{i} x_{ijk}, \sigma^{2}_{i}) \qquad i=1, 2, \ldots, N, \, j= 1,...,n_{ik}, \, k = 0,1
\end{eqnarray}
Here, $\alpha_{i}$ interpreted as the expectation of gene expression values for gene ${i}$ within the control group and $\beta_{i}$ is the expected change in expression of this gene for the case group.  Variances for gene expression are allowed to vary across genes.
The following prior distributions are assumed for $\alpha_{i}$ and $\sigma^{2}_{i}$
$$
p(\sigma^{2}_{i} ) \buildrel {ind} \over \propto  {1/\sigma_i^2} , \qquad
\alpha_{i} \mid  \kappa  \buildrel {ind} \over {\sim }  N(0, \kappa^{2}) \, ;
$$ we take $\kappa$ to be very large.

The traditional hypotheses are $H_i: \, \beta_i =0$ across all genes.  So if all $H_i$ are true, there is no difference in gene expression comparing
healthy to diseased individuals across all genes.  It is here that we can recognize that the problem at hand can be cast in terms of variable selection.  For microarray studies, $N$ is generally quite large and the $n_{ik}$ tend to be small by comparison.  There is also a general belief that most of the null hypotheses will be true.  We are thus expecting a sparsity of non-zero $\beta_i $s.

All that remains is for us to specify a model for the $\beta_i$s.  There are many possibilities in the Bayesian variable selection literature, see for example see \cite{george03} and \cite{ohara09}.  Our model for the regression coefficients is hierarchical where the first level assigns independent normal priors to the $\beta_i$s with distinct variances, namely
\begin{eqnarray}
\beta_{i} \mid \tau_{i}^{2} &  \buildrel {ind} \over {\sim }  & N(0, \tau^{2}_{i})
\end{eqnarray}
  This is termed an adaptive shrinkage prior by \cite{ohara09} (2009).  At the second level, we assume that the first level variances, $\tau_i^2$, are themselves iid from some \emph{unknown} distribution.  At the third level we assume a Dirichlet Process prior for the unknown distribution.  In notation, we have:
\begin{equation}\label{BRDy}
\tau^{2}_{i} \mid G   \buildrel {ind} \over {\sim }  G \qquad
G  \sim  \mathcal{D}(G_0, \gamma)
\end{equation}
where $G_0$ is the expectation of $G$ and $\gamma$ is a weight.  Thus the regression coefficients are modeled as a Dirichlet Process mixture (DPM) of mean zero normal distributions where the mixing is on the variance. Moreover, if $\gamma$ is large then $G$ is approximately $G_0$ while if it is small, the random $G$ allows for realizations that can depart considerably from $G_0$ thus resulting in much greater modeling flexibility. It is well known that the DP is discrete with probability one so this model allows for clustering the variances for the regression coefficients, which induces clustering on the regression coefficients themselves.  Regression coefficients associated with large variances will be presumed to be potentially important.  Our basic model is thus expressed by (2.1-3).

Based on Sethuraman's constructive definition of the DP \citep{sethuraman94}, we have the following representation of the DPM for the regression coefficients
\begin{eqnarray*}
\beta_i \sim  \sum_{c=1}^{\infty}p_{c}N(0, \phi^{2}_{c})
\end{eqnarray*}
 $\phi_c$s are iid from $G_0$, $p_c=b_c\prod_{i=1}^{c-1}(1-b_i)$ where $b_c \buildrel {ind} \over {\sim}$ beta$(1,\gamma)$.  For our problem we choose $G_0$ to be
 Log-$N(m, M^{2})$.  We also assume $\gamma  \sim \textrm{Log-}N(l, L^{2})$.

The distribution of the $\beta_j$s is thus modeled as a countable random mixture of normal distributions with mean zero and different variances.
Thus the $\tau^{2}_{i}$ for different genes may be the same and equal to one of the values of $\phi^{2}_c$. The DPM model naturally creates clusters of genes, where each cluster has its own unique $\phi^{2}$. The number of possible clusters is theoretically infinite \emph{a priori}, but is data driven and there are only a finite number of genes, so there can only be a finite number of clusters. We can rank the identified groups according to their degree of importance using their corresponding $\phi^{2}$. Those $\beta$'s in clusters with $\phi_c^2$ that are near zero correspond to normal distributions with small variances and their corresponding genes must have relatively small regression coefficients and can be considered least relevant. In contrast, those $\beta_i$s that are relatively far from zero would be assigned to normal distributions with larger variances and the corresponding genes considered as more relevant. Therefore, when comparing two subgroups of genes, the one with the highest value of $\phi_c^{2}$ is considered to be most the relevant.  We let $R_{i}$ denote the rank of the $i^{th}$ gene so if $R_i=1$, it is in the cluster with the smallest variance.

\subsection{Inference}
For inference, it is awkward to work directly with the DP so it is common to marginalize over $G$.
We also introduce the latent variable $c_i$, which identifies the cluster to which gene ${i}$ belongs.  Once the magnitudes of
all cluster variances are known, the value of $R_i$ is determined by knowing the value of $c_i$.  Since these values are both latent, we cannot say precisely what they are.  However, since we use Markov chain Monte Carlo (MCMC) numerical methods, and in particular Gibbs Sampling, for approximating the joint posterior distribution, we can obtain samples of these values at each iteration of the Gibbs Sampler.

The actual model specification is greatly simplified by marginalization.  There are a number of different approaches to marginalization in the literature but we follow the one developed by \cite{maceachern98}. First, it is well known that
\begin{eqnarray} 
P(c_j = c \mid c_1, ..., c_{j-1}) & = & \frac{N_{jc} }{j - 1 +\gamma}, \qquad c \in \{c_{1}, \ldots, c_{j-1}\}
\end{eqnarray}
where $N_{jc}$ is the number of genes previously assigned to group $c$. This defines the marginal joint distribution for $(c_1,...,c_N)$.
Then the marginalized model replaces (2.1) and (2.3) with (2.4) and (2.5)
\begin{eqnarray}
y_{ijk} \mid  \alpha_{i}, \{\beta_c \}, c_i & \buildrel {ind} \over { \sim }& N(\alpha_{i}+\beta_{c_i} x_{ijk}, \sigma^{2}_{i}) \nonumber \\
 \phi_c^2 & \buildrel {iid} \over {\sim} & G_0 \quad c = 1,...,K \equiv \hbox{max}\{c_1,...,c_N\}
     \end{eqnarray}
This formulation and posterior inferences for it are well developed in the literature.  From here, it is straight forward to obtain sample the full conditionals for $\phi_c^2\mid \{\phi_j^2: \, j \neq c\}, y$ and for $c_j \mid \{c_k: \, k \neq j\}, y$.  We have used algorithm number 8 in \cite{neal00}, which closely resembles the ``no gaps" algorithm of \cite{maceachern98}.

As previously mentioned, at iteration $p$ of the Gibbs sampler, we are able to ascertain the ranks $R_i^{(p)}$ for $i = 1,...,N$, where $R_i^{(p)}=1$ if $c_i^{(p)} = 1$ and
if $\phi_1^{(p)} = $ min$_c\{\phi_c^{(p)}\} \equiv \phi_0^{(p)}$, and so on.
The number of clusters, $K^{(p)}$, also varies across iterations.  We thus obtain the posterior distribution of these ranks and the number of clusters. We denote the posterior mean of the rank for gene $i$ as $\bar{R}_{i}$ and use it as a measure of relevance. That is, we denote the gene $i$ as relevant if $\bar{R}_{i}$ exceeds a pre-specified cutoff.  Alternatively, we can use the posterior mode (i.e., most frequent rank over $B$ posterior Monte Carlo samples) of the rank $R_{i}$ for each gene as its degree of relevance. We denote this measure of relevance as $\hat{R}_{i}$. As the value of $R_{i}$ increases, the degree of relevance increases.

We also define a relevance measure that is similar to that of GMZ. (See (\ref{eq:v})). To this end, we denote the smallest value of $\phi_c$ at each iteration as $\phi_{0}$. For gene $i$, we create a binary indicator, $s_{i}$, which is set to 1 when $\phi_{c_i} = \phi_{0}$, and zero otherwise. We can use the $B$ posterior Monte Carlo samples to calculate $v_{i} = 1 - \sum_{b=1}^{B} I(s^{(b)}_{i}=1)/B$.  The gene $i$ is then selected as relevant if $v_i$ exceeds a pre-specified cutoff value.  As we will see later, the $v_{i}$ measure obtained from our model could lead to better identification of relevant genes compared to the corresponding $v$ measure based on the corresponding GMZ measure.  We refer to our final model as \emph{BRD (Bayesian Relevance Determination)}.

Relevance measures are used to decide on whether to keep regression coefficients in the model, or not.  Coefficients that correspond to values that exceed a cutoff are kept while those that don't are dropped.  As a method of comparison of different criteria, in section 4 we simulate data from various types of models and obtain a numerical approximation to the true receiver operating characteristic (ROC) curves that correspond to each measure.  The ROC curve plots the proportion of false positive outcomes against the proportion of true positive outcomes.  Here the false positive rate is the proportion of genes that are not differentially expressed for which their regression coefficient was included in the selected model, and the true positive rate is the proportion of genes for which there is a difference between diseased and healthy individuals where the corresponding regression coefficient is left in the model.  The area under the curve (AUC) is a measure of overall performance of the measure as a discriminator among models; if the AUC is one then the correct model is always selected and if it is 0.5, the measure is equivalent to tossing a fair coin e.g. useless.
\subsection{Reduction of the Data}
When the data are reduced to a collection of summary test statistics, our model cannot be modified directly.  When the data are reduced from having two samples (cases and controls) for each gene to a single univariate test statistic, our model (2.1) no longer applies.  Thus, we model the summary statistics, $z_i$, as follows:
\begin{eqnarray}\label{BRDz}
z_{i} \mid  \tau_{i}^{2} & \buildrel {ind} \over {\sim} & N(0, \tau^{2}_{i}) \qquad
 \tau^{2}_{i} \mid  G  \buildrel {ind} \over {\sim}  G \qquad
G  \sim  \mathcal{D}(G_0, \gamma)
\end{eqnarray}
Here, the mean zero for the $z_i$s corresponds to the new $H_i: \, E(z_i) = 0$, e.g. the hypothesis that there is no difference in expression for gene $i$ between diseased and healthy individuals.  This model allows for the partitioning of genes into groups of increasing relevance just as before.  It also allows for the possibility that there is only a single group, which would correspond to no differential gene expression at all.  While details are of course different, inferences for this model proceed in similar way as was discussed for the full data.  Our code for the two models can be found at \url{http://www.ics.uci.edu/~babaks/Site/Codes.html}.

We note that the marginal model for the full data based on (2.1-2) is $$ y_{ijk}\mid x_{ijk} \buildrel {ind} \over {\sim} N(\alpha_i , \sigma_i^2+ x_{ijk }\tau_i^2)\, , \forall i,j,k , $$ while the comparable model here is $  z_{i}  \buildrel {ind} \over {\sim}  N(0, \tau^{2}_{i}) \, , \forall i.$  The remaining parts of both models are identical.  Here, we have no interpretation involving variable selection, but rather model selection.

Note that using summary statistics instead of the full data is not recommended in general. As pointed out by \cite{storey07b} too much information is lost by using gene-level summary statistics such as $z$ scores. However, for illustration purposes, we use summary statistics in some of the simulations discussed in Section \ref{sim} and one of the real datasets in Section \ref{real}.

By focusing on variances, our method has the ability to detect both location shift and scale change in the distribution of $z_i$ (or $\beta$ in the full data case). To see how $\tau^{2}$ can capture location shift, suppose all $z_{i}$ for the relevant group are around 1. Since the mean is fixed at zero, the variance for the corresponding cluster of relevant genes must be large to accommodate these values. Therefore, by fixing the mean at zero, our method has the capability to detect both location shift and scale change.
\subsection{Bayesian Discovery Procedure (BDP)}
The main model proposed and discussed by GMZ (2009), for reduced data, has the following form:
\begin{eqnarray*}
z_{i} \mid  \mu_{i}& \buildrel {ind} \over {\sim} & N(\mu_{i},\sigma^2), \qquad i = 1, \ldots, N \\
\mu_{i} \mid  G  & \buildrel {ind} \over {\sim} & G \\
G & \sim & \mathcal{D}(G_{0}, \gamma) \qquad
G_{0}  = p_{0}h_{\{0\}}(.) + (1 - p_{0})h_{\{0\}^{c}}(.)
\end{eqnarray*}
The baseline distribution $G_{0}$ is a mixture of two terms, one with point mass at zero, and the other with a continuous distribution,
$N(0, \tilde \sigma^{2})$ and technically excluding zero.  $p_0$ is a mixing parameter.  The same model was proposed by \cite{bogdan08}.  A particularly nice feature of this model is that this particular choice of $G_0$ results in a cluster of means all taking the value zero, which corresponds to the collection of null hypotheses being true.  They also allow for point mass at zero to be replaced with a small interval around zero.

They define the indicator $s_{i}$ such that $s_{i} = 1$ when $\mu_{i} = 0$.  Then using $B$ posterior Monte Carlo samples, they use the following measure to set up a threshold in order to divide the genes into ``significant" and ``non-significant" categories:
\begin{eqnarray}\label{eq:v}
v_{i} & = & 1 - \sum_{b=1}^{B} I(s^{(b)}_{i}=1)/B
\end{eqnarray}
GMZ showed that the criterion based on $v_{i}$ can be approximated by the ODP criterion.  They made a number of comparisons with ODP and found that BDP was comparable to better in a number of instances.

The BDP model for reduced data could be employed for relevance determination in the same way that ours is by simply clustering at each iteration of the GS on the distinct values of $|\mu_i^{(b)}|: \, i = 1,...,N$.  Clusters corresponding to the maximum absolute mean are considered most relevant.  Their measure, $v_i$, can also be used to assess what we call relevance in exactly the same way that we use our $v_i$ measure. Using simulated data, we demonstrate in section \ref{sim} that our model for reduced data could provide greater capability, with more parsimonious modeling (e.g., fewer mixture components) to identify relevant genes for further investigation.

GMZ also also adapt their model to handle the full data, $y$ (see the last paragraph of their section 5.2 for details).  They employ a DPM that mixes (and consequently clusters) on $(\mu_{i1},\sigma^{2}_{i})$ ($\mu_{ik}$ is the mean response for gene $i$ in case-control type $k$; $\mu_{i0}$ is forced to zero). They set up their clustering so that $\mu_{i1} =0$ is a possibility through the same device as before.  They create a statistic $v_i$ as before and select genes based on an FDR based cutoff.  Our method based on full data is related to theirs through its use of DPM modeling.  However, their clustering is on the mean-variance combination among the cases while ours is on the effects, $\beta_i$, among the cases.  Our approach amounts to variable selection among these effects and ultimately including or excluding these coefficients in the model based on the magnitudes of their variances, while their approach amounts to using the clustering only through the cluster with point mass of zero.  It is not clear how to use their clusters beyond that here since there is no obvious linear ordering on the two dimensional based clusters.

\section{Data Analysis}
\label{real}

In this section, we apply BRD, BDP, and locFDR to two real data based on two scientific studies.  For the first study, we analyze the observed data directly. For the second study, we model the summary statistics.

We use the ``fdrtool'' package in R to run the locFDR model. The thresholds to divide genes into relevant and not relevant are specified based on $q$-values \citep{storey02}, which are defined as the minimum false discovery rate that is incurred when calling a test significant. (See \cite{storey02} for more discussion.) For the BDP model, we used the R code available online at \url{http://www.math.unm.edu/~michele/Papers/BDP_final_rcode.r}. (In this version of the BDP model, DPM clustering is on both mean and variance.) For our model, we used the log-normal distribution with mean -3 and variance 4 for $G_0$ since this distribution covers the range of feasible values for $\tau^2$ with high probability. The 0.025 and 0.975 quantiles of this distribution are $1.96\times 10^{-05}$ and $1.26 \times 10^{2}$, respectively. We also use $\textrm{Log-}N(-3, 2^2)$ as the prior for the scale parameter $\gamma$. In practice, one could choose a distribution with much narrower 95\% intervals and upper limits closer to zero since the values of $\gamma$ and $\tau^{2}$ tend to be small. We use the posterior mean of rank, $\bar{R}$, to set the thresholds.

\subsection{HCMV Infection}

\begin{figure}[t]
\begin{center}
\includegraphics[height=0.6\textwidth]{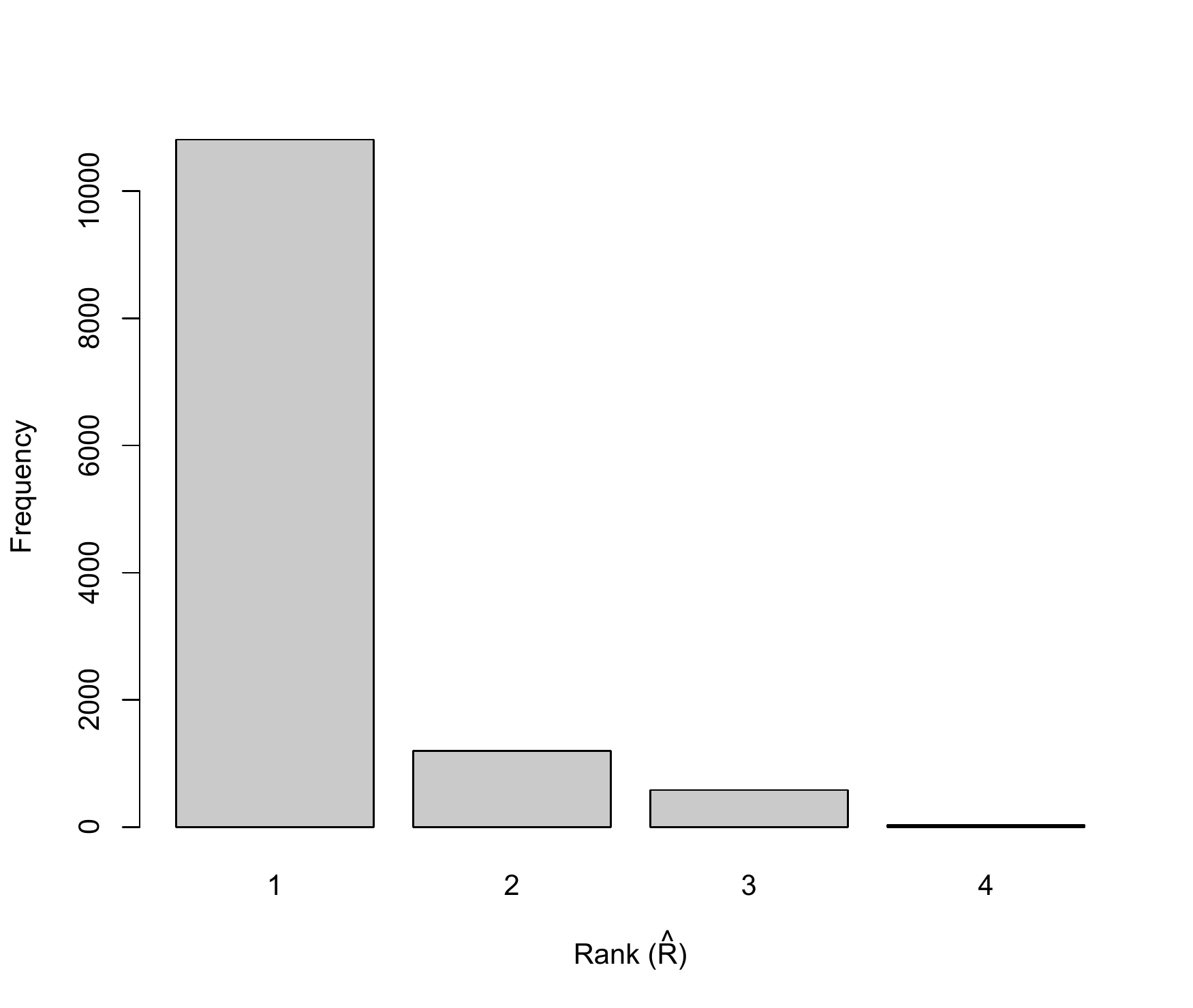}
\caption{Histogram of gene ranks ($\hat{R}$) for the HCMV data. The top ranking group ($\hat{R} = 4$) includes 27 genes.}
\label{hcmvPlot1}
\end{center}
\end{figure}

Our first example involves identifying differentially expressed genes due to infection by human cytomegalovirus. This study was conducted by \cite{chan08}. Out of 12,626 genes, they identified 1,204 genes as statistically and biologically significant. They considered a gene as statistically significant if the observed significance level, $p$-value, was less than 0.05 based on a one-way ANOVA test. Among those genes considered as statistically significant, they only selected biologically significant genes for which at least  four of six HCMV samples changed by 1.5 folds or higher. \cite{chan08} did not adjust their significance cutoff for multiple hypothesis testing. By using local false discovery rate and setting the cutoff $q$-value \citep{storey02} to 0.05, the number of selected genes is reduced to 361. The number of biologically significant genes among these is 328.

For this example, we run our BRD model directly on the observed data, $y$, as opposed to the summary statistics. The bar plot in Figure \ref{hcmvPlot1} shows the distribution of ranks $\hat{R}$. As can be seen, the genes are divided into four groups with respect to their degree of relevance. The most relevant group, $\hat{R} = 4$, includes 27 genes. Of these 27 genes, 18 of them are among the 361 genes selected based on local false discovery rate, and 19 of them are biologically significant (i.e., at least 4 out of 6 samples changed by 1.5 folds). The list of these genes are provided as a supplementary file. Note that some of these genes have relatively high $q$-values and are not considered as relevant based on classical significance tests.

We used the Functional Annotation tool in DAVID Bioinformatics Resources \citep{DAVID} to learn more about the selected genes. We found that 7 of these genes are involved in rheumatoid arthritis, 7 are involved in colorectal cancer, 6 of them in breast cancer, and 5 of them are involved in inflammatory bowel disease. HCMV is in fact known to be related to inflammatory diseases (e.g., rheumatoid arthritis) and several cancers. \cite{naucler08} reviews the evidences for involvement of HCMV microinfections in inflammatory diseases and cancer.

We were also interested in finding the pathways that these genes represent. Using DAVID, we found that 6 of these genes are involved in Toll-like receptor signaling pathway. Further, we identified several Gene Ontology (GO) biological process terms associated with the selected genes. Specifically, we found that 9 of these genes are related to defense response, and 8 are associated with inflammatory response. The relationship between HCMV and these biological processes are well documented \citep[e.g., ][]{booss89, compton03}.

We also used the BDP model for this study. The minimum value of $v$ over all genes was 0.998, which means no gene was selected as significant.

As discussed above, we could also use $z$ scores instead of the full data. For our model, this leads to the selection of 23 genes that correspond to $\hat{R} = 5$; genes were divided into 5 groups based on their relevance. Six of these genes are among the top 27 that were selected by our model using the full data. For the BDP model, 631 genes were selected using the cutoff 0.01 for $v$. Among the top 23 genes selected using BDP, only 3 appear in the list of 23 genes based on BRD. Thus the methods can give quite distinct results.

\subsection{Leukemia}
Our second example, which is a well-known study typically used as benchmark problem, involves identifying differentially expressed genes between two types of leukemia: acute myeloid leukemia (AML) and acute lymphoid leukemia (ALL). This study was conducted by \cite{armstrong02}. The data include the expression levels of 10,056 genes for 48 subjects (24 subjects in each group).
\begin{figure}[t]
\begin{center}
\includegraphics[height=0.4\textwidth]{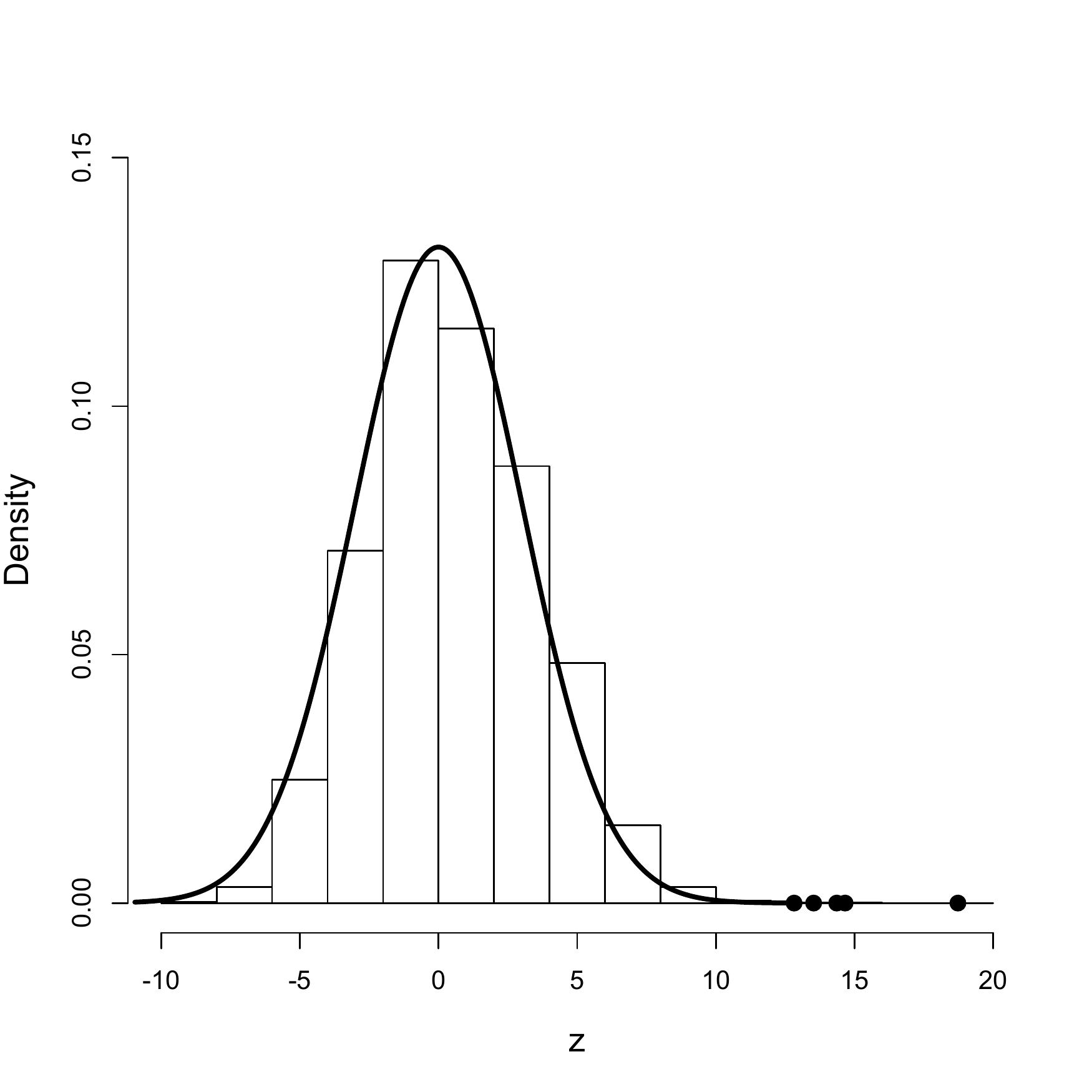}
\caption{Histogram of observed $z$ scores for the leukemia data. The superimposed curve is the predictive density based on $\phi^{2}_{0}$, which is the smallest value of $\phi^{2}$ at each iteration. The five points identified on the plot are the $z$ scores for the genes that are identified as relevant based on $R$.}
\label{fig:leuk}
\end{center}
\end{figure}

The histogram of $z$ scores for the leukemia data is shown in Figure \ref{fig:leuk}. For our model (\ref{BRDz}), we ran 4000 iterations of MCMC, discarded the first 1000 iterations, and used the remaining iterations to obtain posterior distributions. Convergence of the chain was verified based on the trace plots of hyperparameters. As mentioned above, at each iteration, we denote the smallest value of $\phi$ as $\phi_{0}$. The curve in Figure \ref{fig:leuk} shows the predictive density based on $\phi_{0}$. To obtain this curve, we found the density function for $N(0, \phi_{0}^{(b)})$ at iteration $b$, and averaged these over MCMC iterations.

The posterior mode of rank, $\hat{R}$, is 1 for most genes. However, there are five genes whose posterior mode of rank is $\hat{R}=2$. These genes (in the descending order of $\bar{R}$) are TCL1A, DNTT, CD24, TOP2B, and PSMA6. The values of $v$ for these genes are 0.011, 0.048, 0.06, 0.107, and 0.19 respectively. These genes are represented by individual data points in Figure \ref{fig:leuk}.

The identified genes all are known to be associated with leukemia. Specifically, TCL1A (T-cell leukemia/lymphoma 1A) is known to be upregulated in ALL patients \citep{zangrando09}. DNTT is also upregulated in B lineage ALL compared to AML \citep{farahat95}. In their paper, \cite{raife94} showed that expression of CD24 predicts monocytic lineage in AML. The resistance of several leukemia cell lines to therapy has been associated with a decreased protein expression and/or activity of TOP2B (topoisomerase II) enzymes \citep{deffie89}. Lastly, while the effect of PSMA6 on ALL vs. AML is not well studied, \cite{chen09} have recently shown that the expression levels of PSMA6 in AML-M5 leukemia cells was low compared to AML-M5 leukemia cells and normal blood cells.

We also used the BDP model for analyzing the leukemia data, and ranked the genes based on their values of ${v}$. The five genes identified by our model are also the top ranking genes using BDP.

Using locFDR and setting the cutoff for $q$-value at 0.05, we select 2652 genes as significant. The top 5 genes based on this method are TOP2B, TCL1A, PSMA6, CD24, and CD79A. Four of these genes also appear among the top five genes based on BDP and BRD.

\section{Simulations}
\label{sim}
In this section, simulated summary data are used to compare the results of locFDR, BDP and BRD. We also use the full data to compare our method to BODP \citep{cao09} as well as locFDR and BDP. We compare different methods using the area under the receiver operating characteristic (ROC) curve (AUC) for identifying relevant genes.

Our first simulation study, Simulation 1, is similar to that of GMZ (2009). We generate 500 $z$ scores. (Note that we simulate $z$ scores as opposed to gene expression values $y$ for simplicity.) The first 40 $z$ scores are sampled from $N(\mu_{k}, \sigma^{2})$, where $\sigma^{2}=1$ and $\mu_{k} = -1, 1, 2, 3$ with equal proportions. The remaining 460 are sampled from $N(0, 1)$. All models are expected to identify the first 40 $z$ scores as relevant. Note that the data are following the BDP model very closely (i.e., mixture components have different means and the same variance) so we expect it to outperform the BRD model.

\begin{figure}
\begin{center}
\includegraphics[height=0.4\textwidth]{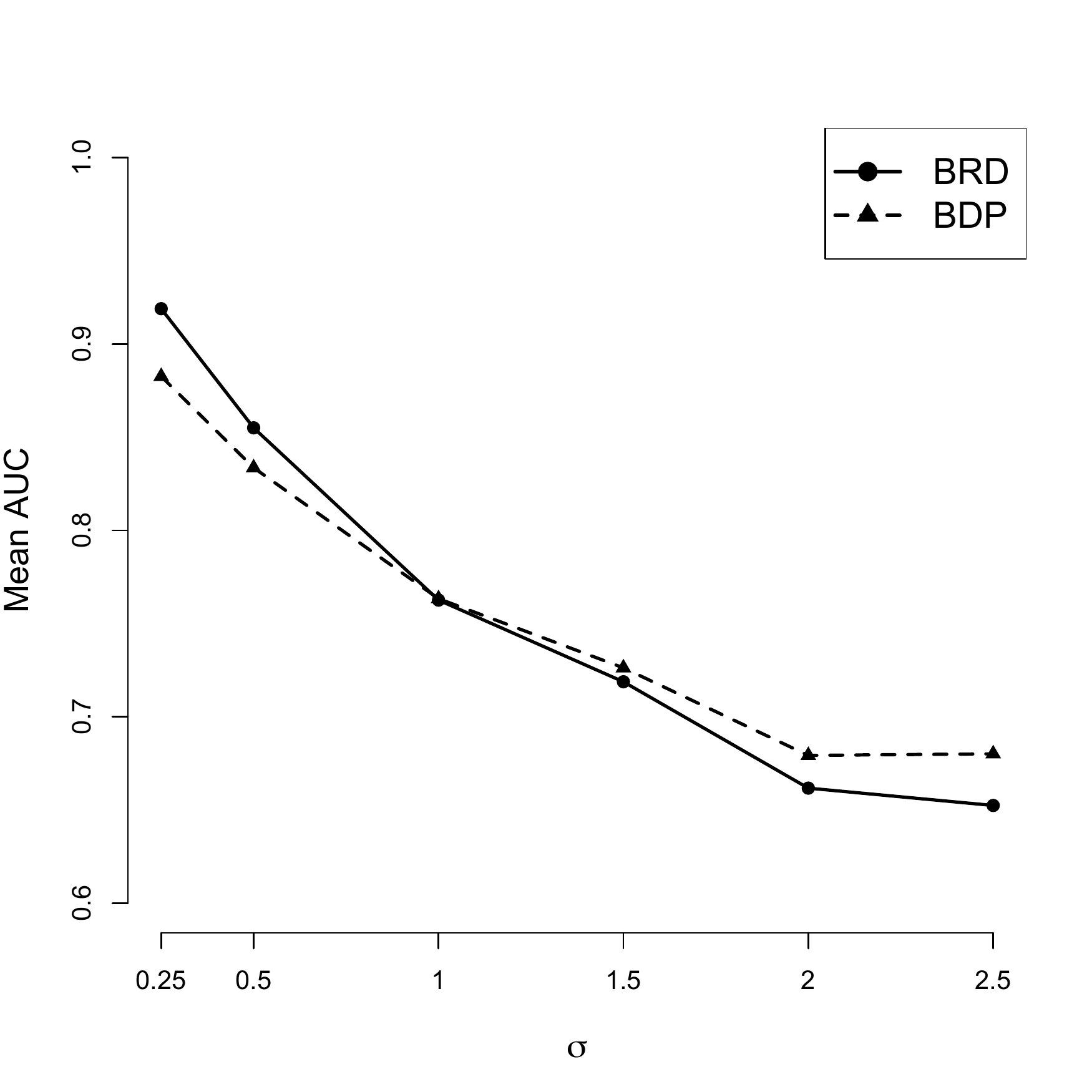}
\caption{Increasing the standard deviation, $\sigma$, in Simulation 1 gradually from 0.25 to 2. The data are generated according to the BDP model. While BRD performs better than BDP initially, the BDP model outperforms our model for higher values of standard deviation.}
\label{fig:power}
\end{center}
\end{figure}

Our criterion for comparison of methods is to find AUCs, which are used to determine how well diagnostic procedures discriminate between groups.  Table \ref{dpSim1} shows the average AUC using 100 simulated data sets. The corresponding standard errors (SE) are presented in parentheses. We see that BRD has a statistically higher average AUC than BDP ($p$-value $< 0.01$ using a paired $t$-test), however the difference is not substantially large. Since outcomes depend on the specific conditions under which $z$ scores are generated, we investigated further, repeating the above simulation with different values of $\sigma$. For the first 20 data sets, we set $\sigma = 0.25$. Under this scenario, the relevant genes are easier to identify. For the second 20 data sets, we increased $\sigma$ to 0.5. This makes the identification of relevant genes slightly difficult. We continued increasing $\sigma$ to 1, 1.5, 2 and 2.5 for each consecutive set of 20 simulated data sets. Figure \ref{fig:power} compares the two models, BDP and BRD, in terms of average AUC for each value of $\sigma$. While, BRD performs slightly better than BDP for small values of $\sigma$, BDP performs better than BRD for relatively larger values of $\sigma$, where identifying relevant genes becomes difficult. This was of course expected since the data matches the BDP model more closely. In what follows, we consider alternative scenarios where data are not generated according to any of the above three models.

For our second simulation study, Simulation 2, we allow the distribution of the $z$ scores for not relevant genes to deviate from normality. To this end, we sample 460 $p$-values for these genes from Beta(6, 4) and obtain their $z$ scores by applying the inverse cdf of the standard normal distribution to their $p$-values. For Beta(6, 4), the 95\% interval is $(0.30, 0.86)$. While these $p$-values are generally considered as ``non-significant," they are not uniformly distributed between 0 and 1. Therefore, the resulting distribution of $z$'s is not $N(0, 1)$. For the relevant genes, we sample between 5 to 10 $p$-values from Beta(6, 40), whose 95\% interval is $(0.05, 0.24)$. As before, to obtain the $z$ scores for these relevant and moderately relevant genes, we apply the inverse cdf of the standard normal distribution to their $p$-values. Table \ref{dpSim1} shows the average (over 100 data sets) AUC and the corresponding standard errors for the three models. This time, our model performs substantially better than the two alternative models.

For our third simulation study, we try to make the simulated data as realistic as possible. To this end, we use the actual gene expression values and the class labels from real biological data obtained based on interrogating the mutation status of p53 in cancer cell lines. The data are publicly available from \url{http://www.broadinstitute.org/gsea/datasets.jsp}. Out of $n=50$ cell lines, 17 were classified as normal and the remaining 33 were classified as muted. The data include 9,703 human cDNAs, which correspond to approximately 8,000 different genes. For each gene, we calculate the $z$-score using the $t$-test based on the difference in the gene expression between normal and mutated cells. We use these $z$-scores (approximately 8,000) to simulate 100 data sets. For this, we sort the $z$ scores according to the their absolute values. Then, we randomly sample between 40 to 80 $z$ values from the top 400 genes (i.e., genes with highest values of $|z|$), and sample 460 $z$ values from low ranking genes (i.e., genes with relatively lower values of $|z|$). A good model is expected to regard the first set of genes as relevant and the remaining 460 genes as not relevant. As before, we use AUC to compare the above three methods. Our model outperforms the other two methods by a substantial amount (Table \ref{dpSim1}).

Simulation 4 is similar to Simulation 3 (i.e., we use the actual $z$ scores from p53 data) but this time we create two groups of relevant genes with different degrees of relevance. To this end, we randomly select 5 to 10 $z$ scores from the top 100 genes and regard them as relevant. Then, we remove the top 200 genes from the data. From the remaining genes, we select 20 to 30 of top ranking $z$ scores and regard the corresponding genes as moderately relevant. Finally, we randomly sample 460 of low ranking $z$ scores for the not relevant group. This time, BDP performs slightly better than our model if we use ${v}$ as a measure of relevance (Table \ref{dpSim1}). The difference, however, is not statistically significant ($p$-value = 0.23 using a paired $t$-test). If we use the posterior mean of rank, $\bar{R}$, as a measure of relevance, our model performs substantially better than BDP ($p$-value = 0.02 using a paired $t$-test).

\begin{table}
\centering
\caption{\label{dpSim1} Comparing BRD to BODP, locFDR, and BDP based on the area under the ROC curve (AUC). The corresponding standard errors are shown in parentheses.}
\begin{tabular}{l  c c  c c c }
AUC\% & locFDR  &  BODP & BDP & \multicolumn{2}{c}{BRD} \\
\cline{5-6}

\vspace{-6pt}
& & &  &  \\

& & & & $v$ & $\bar{R}$ \\
\hline
Simulation 1 & 75.8 (0.7) & - &  75.8 (0.4) & 76.8 (0.4) & 77.0 (0.4) \\
Simulation 2 & 90.9 (1.5) & - & 88.9 (0.4) & 93.5 (0.3) & 94.8 (0.2) \\
Simulation 3 & 64.5 (2.3) & - & 82.2 (1.3) & 90.3 (0.7) & 94.4 (0.4)\\
Simulation 4 & 57.9 (1.8) & - & 88.5 (1.1) & 86.8 (0.8) & 91.8 (0.6) \\
Simulation 5 & 73.7 (1.1) & 76.2 (1.2) & 65.1 (1.0) & 79.6 (1.0) & 77.7  (1.0)\\
Simulation 6 & 75.5 (1.1) & 72.8 (1.2) & 63.9 (1.0) & 77.6 (1.1) & 78.3  (1.1)
\end{tabular}
\end{table}

So far, our simulations have been based on summary statistics. Next, we evaluate the performance of our method based on the full data  as discussed in Section \ref{fullData}. For Simulation 5, we randomly sample 10 cell lines and 250 genes from the p53 dataset. We permute the labels (normal vs. muted) of these cell lines. In this way, all 250 genes would become irrelevant with respect to the new labels, denoted as $\tilde{x}_{ijk}$, where $i = 1, \ldots, 250$, $j = 1, \ldots, 20$, and $k = 0, 1$. Here, $k=0$ denotes the control group and $k=1$ the case group after permutation of the labels. For the first five genes (i.e., $i=1, \ldots, 5$), we add a constant to the gene expression values of the case group (i.e., $k=1$). The constant is set to -1 with the probability of 0.7 and to 2 with the probability of 0.3. Therefore, the first five genes in this simulation are considered as relevant, while the remaining 245 genes are irrelevant. Results for this simulation are shown in Table \ref{dpSim1}. As before, our method performs better than BDP and locFDR. It also outperforms BODP. However, the improvement is statistically significant (at the 0.05 level using a paired $t$-test) based on $v$ only; the improvement based on $\bar{R}$ is marginally significant ($p$-value = 0.07). For the BDP model, in consultation with the authors, we used an Inv-$\gamma(1, 1)$ as the prior distribution of $\sigma_{i}^{2}$ and set $k_{0} = 2$. (See Equation 16 in GMZ.). For this model, the low value for AUC is unexpected. While we did perform other simulations with similar results, we recommend caution about generalizing any of these empirical results. Our method was designed based on variable selection as a goal, and simulation 5 has indeed created a variable selection problem, so perhaps this is what gives BRD the advantage. If we calculate the the $z$ scores for simulated data and run BDP and BRD based on summary statistics, AUC's increase to 94.6, 97.4, and 97.5 for BDP, BRD using $v$, and BRD using $\bar{R}$ respectively. Both models perform substantially better than locFDR. 

For our sixth simulation, we follow a similar procedure as in Simulation 5, but we increase the sample size to 20 and sample, from the standard normal distribution, the values that are added to the gene expression of the case group. As before, our model provides the highest AUC (Table \ref{dpSim1}), and the improvement is statistically significant compared to all other methods. Also, similar to Simulation 5, AUC is unexpectedly low for BDP. However, if we use the $z$ scores, as opposed to full data, for this model, AUC increases to 70.0. For our model, using $z$ scores reduces AUC slightly to 75.1 and 75.2 based on $v$ and $\bar{R}$ respectively. 

Finally, we compare the two nonparametric models, BDP and BRD, when no gene is related to the outcome of interest and all observed effects are due to chance alone. In this case, we expect the two models to assign all genes to the one group, which will be regarded as not relevant; that is, the mixture distribution should have one component only. To evaluate the two models, we simulated 500 $z$ scores from the standard normal distribution. We ran both models on the data and obtained the \emph{mode} of the number of the mixture components over MCMC iterations. We denote this measure as ${C_{m}}$. For the BDP model, the average of $C_{m}$ over 100 data sets was 2.97 (SE=0.07). The average of $C_{m}$ for the BRD model was 1.03 (SE = 0.017), which is substantially lower than that of BDP.

\begin{figure}[t]
\begin{center}
\includegraphics[height=0.4\textwidth]{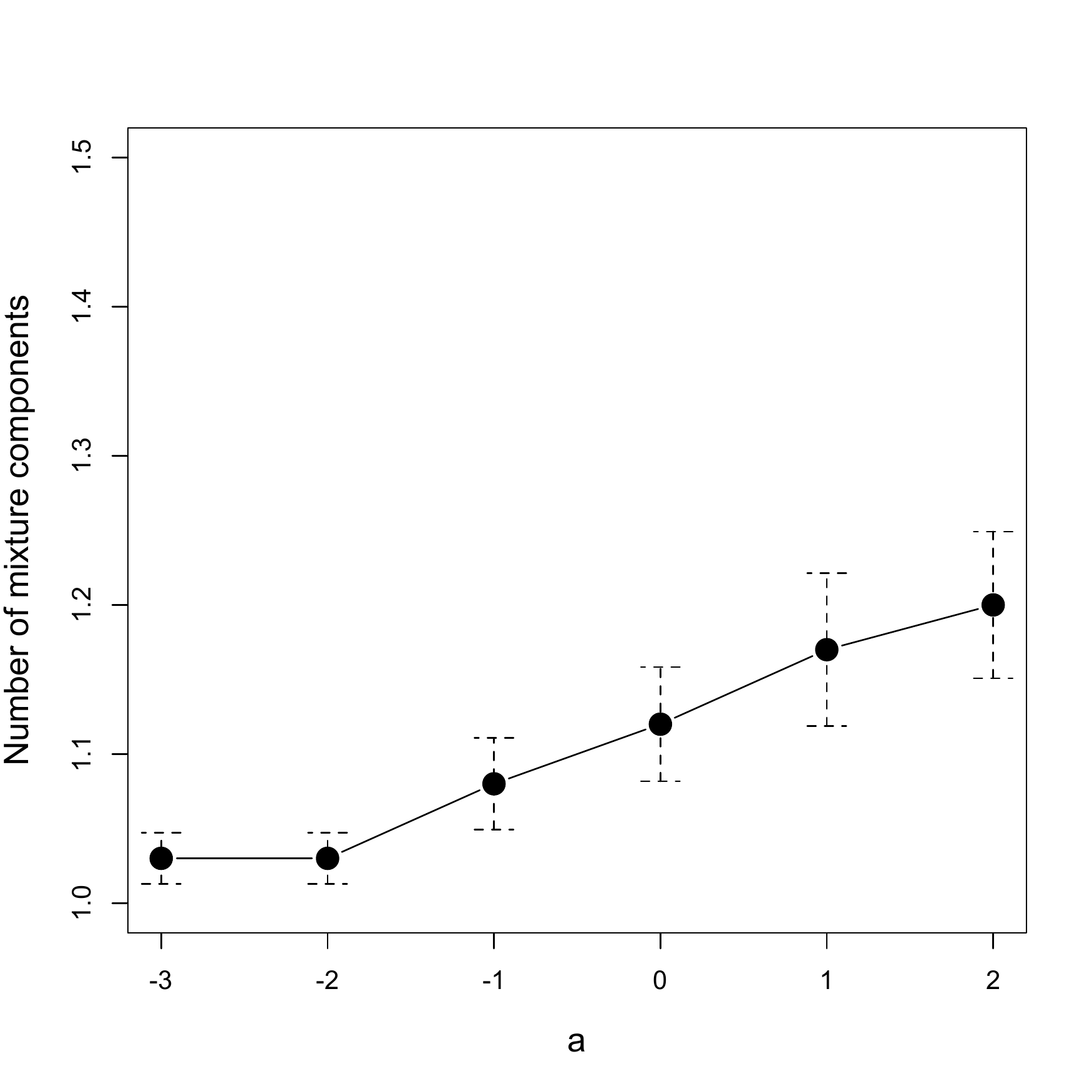}
\caption[]{Average number of mixture components for different values of $a$ when no gene is relevant to the disease. Here, $\log{\gamma}\sim N(a, 4)$. The vertical lines show the corresponding confidence interval for each mean. Note that as $a$ increases, the prior puts higher probabilities on large number of components.}
\label{fig:sens}
\end{center}
\end{figure}

\section{Sensitivity Analysis}

A key parameter in our model is $\gamma$, which is a positive scale parameter that controls the number of components of the mixture that will be represented in the sample. A larger value of $\gamma$ results in a larger number of components. In the previous section, we showed that our model performs reasonably well when there is no relevant gene in the data by keeping the number of mixture components close to one. In this section, we investigate the influence of the specified prior for $\gamma$ on the number of identified mixture components, $C_{m}$, where no gene is relevant.

In our model, we assumed the following prior distribution for $\gamma$:
\begin{eqnarray*}
\gamma & \sim & \textrm{Log-}N(a, b^{2})
\end{eqnarray*}
To change this prior, we set $a$ to -3, -2, -1, 0, 1, and 2 while fixing $b$ at 2. By increasing $a$, the prior puts more probability on large values of $\gamma$, and in turn increases the prior probability of having large number of components.

We simulated 100 data sets for each value of $a$. Figure \ref{fig:sens} shows that the average of $C_{m}$ increases from 1.03 to 1.2 as we increase $a$ from -3 to 2. Note that for most practical problems, values such as 1 and 2 for $a$ lead to unreasonably wide priors; for example, the 95\% interval is $(0.05, 136)$ with $a=1$. In practice, values such as $a=-3$ or $a=-2$ seem more reasonable; The resulting priors based on these values put almost all the prior probability on values of $\gamma$ between 0 and 10.

\section{Discussion}
\label{discussion}
We have proposed a new approach for analyzing large-scale studies, where the objective is to identify factors that are relevant to an outcome of interest. Our proposed method divides a set of candidate factors (e.g., genes) into several subgroups according to their degree of relevance. Not only does our method provide a flexible model for analyzing high throughput studies, but also it simplifies the task of selecting a subset of factors for possible follow-up studies.

Our method uses the Dirichlet process to introduce a random grouping on factors (e.g., genes). It is common to refer to such priors as ``Polya Urn'' to accentuate the clustering aspect of the prior. Many alternative priors have been recently introduced in the literature. For example, the normalized inverse Gaussian process \citep{lijoi05} allows different priors on the clustering while keeping the posterior inference simple.

While we have applied our model to the analysis of gene expression microarrays, our approach can be applied to a wide range of problems, where the relevance of many factors are simultaneously investigated. For example, functional neuroimaging is used to study normal vs. pathological brain processes. A typical experiment in this area involves assessing a large number of pixels, each representing a small area of brain tissue. Another possible application for our method is the analysis of single-nucleotide polymorphisms in genome-wide association studies, whose objective is to identify and characterize genetic variants related to common complex diseases.

In this paper, we presented our model for a binary health outcome variable and continuous responses. Our model could easily be extended to multiple health outcomes and discrete responses. This is especially simple if we use summary statistics.

The main challenge of our model is its high computational cost. For the leukemia data, running our code takes 2.92 seconds of CPU time per iteration. While this is much smaller compared to the BDP model, which takes 32.75 seconds CPU time per iteration (perhaps an unfair comparison since the two codes were developed independently and without intention for comparison), it is possible to reduce the computational cost by using more efficient MCMC algorithms. For example, we could apply the ``split-merge'' approach, which follows a Metropolis-Hastings procedure that resamples clusters of observations simultaneously rather than incrementally assigning a single observation \cite{jain07}. Alternatively, one could use the method of \cite{dahl09} to find the maximum \emph{a posterior} (MAP) estimates (instead of the posterior distribution) of cluster structure in $1-d$ Dirichlet process mixtures.

Future directions could involve incorporating additional knowledge about the underlying structure of data. For example, we could incorporate prior information on the interconnectivity among genes. In this way, knowing that a gene, $\mathcal{G}_{j}$, is differentially expressed could increase the probability of differential expression for other genes that are related to that gene (i.e., either they activate $\mathcal{G}_{j}$ or are activated by it). To this end, we can use relevant biological data to group genes into subsets of related genes and shift the focus of analysis towards gene sets as opposed to individual genes \citep{shahbabaBGSA11}. 

Another possible research direction involves extending our model to allow for incorporating more information on subjects. For example, we could include clinical measures and demographic variables in our model.

Finally, there is much scope for theoretical investigation of the DPM based approaches.  We are currently developing a first attempt at this investigation and expect to continue our investigation in future work.

\bibliographystyle{imsart-nameyear}

\end{document}